\documentclass[aps,prd,preprintnumbers]{revtex4}
\usepackage{graphics}
\def \beq{\begin{equation}}
\def \eeq{\end{equation}}
\def \beqa{\begin{eqnarray}}
\def \eeqa{\end{eqnarray}}

\def\ie{{\sl i.e.\/}}
\def\D{{\cal D}}
\def\P{{\overline P}}
\def\t{\vartheta}
\def\x{{\mathbf x}}
\def\etal{{\sl et al.\/}}

\def\jhep{{\sl J.\ H.\ E.\ P.\/}}

\def\np{{\sl Nucl.\ Phys.\/}}

\def\pr{{\sl Phys.\ Rev.\/}}
\def\prl{{\sl Phys.\ Rev.\ Lett.\/}}

\begin{document}
\title{Relativistic diffusion and heavy-ion collisions}
\author{Rajeev S.\ \surname{Bhalerao}}
\email{bhalerao@tifr.res.in}
\affiliation{Department of Theoretical Physics, Tata Institute of Fundamental
         Research,\\ Homi Bhabha Road, Mumbai 400005, India.}
\author{Sourendu \surname{Gupta}}
\email{sgupta@tifr.res.in}
\affiliation{Department of Theoretical Physics, Tata Institute of Fundamental
         Research,\\ Homi Bhabha Road, Mumbai 400005, India.}

%\pacs{25.75.-q, 24.10.Nz, 25.75.Ld, 12.38.Mh}
\preprint{TIFR/TH/09-05}

\begin{abstract}
We study first- and second-order theories of relativistic diffusion coupled
to hydrodynamics under the approximation, valid at mid-rapidity in the
RHIC and LHC, that conserved number densities are much smaller than
the entropy density. We identify experimentally accessible quantities
of interest, and show that the first- and second-order theories may lead
to radically different evolutions of these quantities. In the first-order
theory the memory of the initial state is almost completely washed out,
whereas in the second order theory it is possible that freezeout occurs
at a time when transient dynamics is still on, and the memory of the
initial state remains. There are observational consequences which we
touch upon. In the first-order theory, and for initial conditions when
the second-order theory mimics the first-order, one may be able
to put a bound on the diffusion constant.
\end{abstract}
\maketitle

Event-to-event fluctuations of conserved quantum numbers in heavy-ion
collisions have been espoused as signals of the underlying thermodynamics
\cite{fluct}. Quantities which have been considered interesting
include the baryon number, $B$, the electric charge, $Q$, and even the
strangeness, $S$, which is conserved in strong interactions. The proton
number, $N_p$, has been suggested as a proxy for $B$ \cite{hatta}. Of
interest are the net conserved quantum number and its distribution over
the ensemble of events at the collider.

In every observation of a collision event, one can find the density
of $B$, $Q$ or $S$ as a function of rapidity. In the usual measurement of
fluctuations one usually sums such a density over all rapidity within the
observational window: this corresponds to taking the Fourier coefficient
$k=0$ of the density profile. However, other Fourier coefficients can be
easily constructed. In this paper we point out that hydrodynamic evolution
of the Fourier coefficients of conserved number densities are interesting
in their own right, since they may contain interesting signals not only
of the initial conditions but also of hydrodynamic evolution.

To this end we investigate the coupled evolution of the number
densities and the usual hydrodynamic quantities, \ie, the stress tensor,
$T_{\mu\nu}$, expressed as usual in terms of the field of flow velocity,
$u_\mu$, and the energy density, $\epsilon$, or the entropy density,
$s$, and the dissipative parts.  In heavy-ion collisions at the
RHIC net conserved number densities are small compared to the entropy
density. This is usually expressed as the observation that the ratio
of the appropriate chemical potential, $\mu$, and the temperature,
$T$, \ie, $\mu/T$, is small \cite{cleymans}. This ratio is expected to
become even smaller at the forthcoming LHC experiment. As a result, the
finite $\mu$ corrections to the pressure, $p$, the speed of sound, $c_s$,
and to $\epsilon$ and $s$ are expected to be small.  The combination of
these quantities along with the number densities obey a set of coupled
equations.  The smallness of $\mu/T$ implies that the equations can be
linearized in $\mu/T$, and an accurate picture of the hydrodynamics is
obtained by considering the diffusion equation coupled to the hydrodynamic
flow while neglecting the effect of the number densities on the flow. We
note that this approximation is made here for convenience. It allows us
to understand the essential physics of the situation without much of
a sacrifice in accuracy. If higher accuracy is needed, one can easily
investigate the fully coupled formalism \cite{coupled}.

Previous work had investigated heavy quark diffusion in the heavy-ion
induced fireball \cite{hqd}. The energy loss of the heavy-quark
was estimated using weak-coupling theory \cite{weak}, the Boltzmann
equation \cite{moore}, AdS/CFT techniques \cite{ads}, and using Langevin
dynamics \cite{hatsuda}.  A diffusive theory has also been invoked to
describe stopping and early time entropy production \cite{wolschin}. Our
motivation, and therefore, the theory we develop, is quite different.

In this paper we consider only longitudinal background fluid flow,
\ie, flow in which the $z$-axis is defined by the direction of the
incoming nuclei, and the dependence of all quantities on the orthogonal
$x$ and $y$ coordinates is neglected. This first approximation is
expected to be valid in nearly head-on (small impact parameter)
collisions of nuclei for times shorter than the sound travel time
across the fireball, \ie, for $\tau\le R/c_s$, where $R$ is the radius
of the colliding nuclei. We work with curvilinear coordinates, \ie,
the space-time rapidity, $\eta=(1/2)\log(t+z)/(t-z)$, and the epoch,
$\tau=\sqrt{t^2-z^2}$. The covariant derivatives, $d_\mu$ then include
Christoffel symbols.  We define the local timelike derivative,
$D=u^\mu d_\mu$. This allows us to define a vector $v_\mu=(D
u_\mu)/S$ which is a spacelike unit vector orthogonal to $u_\mu$,
where $S^2=-(Du_\mu)(Du^\mu)$. In terms of this new vector we define the
spacelike derivative ${\widetilde D}=v^\mu d_\mu$. The spacelike projector
is $\Delta_{\mu\nu}=g_{\mu\nu}-u_\mu u_\nu$. A scalar which appears often
is $\Theta=d_\mu u^\mu$.  For longitudinal flow we can parametrize the
fluid velocity as $u_\tau=\cosh y$ and $\tau u_\eta=\sinh y$ with the
other two components vanishing. Boost invariant flow corresponds to
taking $y=0$ everywhere.  This is not the same as the fluid being at
rest; the choice of the particular curvilinear coordinates that we use
makes this simple parametrization possible. Then one has $\Theta=1/\tau$
and $S=0$ (note that $v_\mu$ remains finite). The derivative $D=\partial
/\partial\tau$ and $\tau{\widetilde D}=\partial /\partial\eta$.  In all
this we follow the notation of \cite{hicus}, to which we refer readers
for further details.

We examine the kinematics of the number current, $n_\mu$. One finds
the number density as usual, $n=n_\mu u^\mu$. Then one can write
$n_\mu=n u_\mu + \nu v_\mu$, where $\nu v_\mu$ is the dissipative part of
the number current. Clearly, one has $\nu=-v_\mu n^\mu$. When there is
more than one conserved particle number, one can write an analogous
decomposition for each conserved quantity, with different $n$ and $\nu$
for each. For each conserved number, the equation of continuity is
\beq
   0 = d_\mu n^\mu = Dn + n\Theta + {\widetilde D}\nu + S\nu.
\label{conti}\eeq
We will assume in this paper that the background fluid undergoes boost
invariant longitudinal flow. However, as a result of the assumption
that $\mu/T\ll1$, there is no contradiction in assuming that $n_\mu$ is
not boost invariant while the background flow is. In fact, the physics
of diffusion, which is the central object of interest here, would be
trivially absent if $n$ and $\nu$ were both independent of $\eta$.

In ideal fluids there is no dissipative part to any hydrodynamic
quantity, so $\nu=0$. The hydrodynamics then lies entirely in the
continuity equation and its self-similar solution
\beq
   \frac{dn(\tau,\eta)}{d\tau} = -\frac{n(\tau,\eta)}\tau,
      \qquad{\rm so\ that}\qquad
   \frac{n(\tau,\eta)}{n(\tau_0,\eta)} = \frac{\tau_0}\tau.
\label{idealsol}\eeq
Any initial number density profile is then propagated in time unchanged
in shape but attenuated as $1/\tau$ due to the geometry of longitudinal
flow. We call this phenomenon Bjorken attenuation. 

Following \cite{hicus} we define the Fourier transform and the power
spectrum of the density, respectively, as
\beq
   n(\tau,k) = \frac1{\sqrt{2\pi}}\int d\eta n(\tau,\eta)\exp[-ik\eta],
   \quad{\rm and}\quad
   P(\tau,k) = |n(\tau,k)|^2.
\label{fourier}\eeq
Note that the measure used here is $d\eta$, whereas the invariant
volume measure contains $\tau d\eta$. Hence, the Fourier coefficient
$n(\tau,k=0)$
is not the conserved charge, but $\tau n(\tau,k=0)$ is conserved. For an
ideal fluid all Fourier coefficients of the number density profile also
undergo Bjorken attenuation, which, as we see, is a purely geometric
phenomenon for boost invariant flows.

For non-ideal fluids
the first order constitutive equation for diffusion is Fick's Law. In
the local rest frame of the fluid this takes the form
\beq
   0= n^i + {\D} \partial^i n,
\label{fick}\eeq
where ${\D}$ is the diffusion coefficient. In a general frame one can
recast this in the form
\beq
   0 = \Delta^{\mu\nu}n_\nu - {\D}\Delta^{\mu\nu} d_\nu n
     = \nu v^\mu - {\D}\left[d^\mu n - u^\mu Dn\right],
\label{relfick}\eeq
using the projector orthogonal to $u$. When the last form on the right
is contracted with $u^\mu$, one finds that it vanishes trivially. A
contraction with $v^\mu$ relates the dissipative part with the spatial
derivative of the number density---
\beq
   0 = -\nu - {\D}{\widetilde D}n.
\label{scalfick}\eeq

Substituting this form of Fick's law into the continuity equation and
thereby eliminating $\nu$, one finds the relativistic version of the
diffusion equation in a fluid undergoing longitudinal expansion---
\beq
   0 = Dn - {\widetilde D}{\D}{\widetilde D}n -
    {\D}S{\widetilde D}n + n\Theta.
\label{diffueq}\eeq
Note that the last two terms are directly related to the hydrodynamic flow.
Since the equation is linear, it can be solved by Fourier transforming in
the spatial variable and considering the evolution of each Fourier mode
separately. For later convenience we give the name Fick diffusion to
phenomena that arise from this equation.

The relative importance of flow and continuity versus diffusion can be
quantified through the dimensionless quantity which compares the 2nd and
the last terms in eq.\ (\ref{diffueq})---
\beq
   {\cal W} = \frac{\lambda^2}{\tau{\D}} = \frac\tau{\D}\sinh^2\Delta\eta,
\label{wiener}\eeq
where $\lambda=\tau\sinh\Delta\eta$ is an intrinsic length scale in the
density profile, corresponding to a scale $\Delta\eta$ in rapidity. When
${\cal W}\ll1$ then diffusive effects dominate over flow; when it is much
greater than unity, flow dominates. Which behaviour dominates depends
also on how $\D$ changes with time. In the high temperature phase of
the plasma, where there is essentially only one scale, $T$, one has
$\D\propto1/T$ (in an AdS/CFT computation one typically obtains $\D T =
1/2\pi$, although much smaller and larger results can also be obtained by
tuning parameters \cite{einstein}).  However, at sufficiently small $T$
(close to $T_c$, for example) it is possible that $\D$ is controlled by
a hadronic length scale, and may change only marginally with $T$. In the
high temperature phase, if the background flow is boost invariant and
longitudinal, then $T\propto1/\tau^{(1+c_s^2)/4}$. In this case $\cal W$
increases with $\tau$, leading to a decreased importance of the diffusive
term. At lower temperature as well, since $\D$ varies little, $\cal W$
increases and the flow terms become more important. If we are concerned
with a $\Delta\eta$ range such that diffusion dominates at initial times,
then after a time $\tau_{fl}=\D/\sinh^2\Delta\eta$, flow dominates over
diffusion. If at times less than $\tau_{fl}$ diffusion manages to destroy
structures in the number density profile, then the memory of the initial
state can be lost.

This can be seen in a simple model of a fluid where all transport
coefficients are constant and independent of parameters such as $\mu$
and $T$. Then, for any longitudinal background flow of
such a fluid the diffusion equation (\ref{diffueq}) becomes
\beq
   0 = \frac{\partial n}{\partial\tau} + \frac n\tau
       - \frac{\D}{\tau^2}\frac{\partial^2n}{\partial\eta^2},
\label{simplediffu}\eeq
after linearizing in $\mu/T$. The Fourier coefficients of $n$ obey the
equation
\beq
   0 = \frac{\partial n(\tau,k)}{\partial\tau} + \frac{n(\tau,k)}\tau
       + \frac{{\D}k^2}{\tau^2}n(\tau,k).
\label{fourdiff}\eeq
The solution can be obtained by quadrature---
\beq
   n(\tau,k) = n(\tau_0,k) \left(\frac{\tau_0}\tau\right)
    \exp\left[-\frac{{\D}k^2}{\tau_0}\left(1-\frac{\tau_0}\tau\right)\right].
\label{solvdiff}\eeq
The constant mode ($k=0$) exhibits Bjorken attenuation, as all
Fourier components eventually do, consistent with the analysis of
$\cal W$. Behaviour typical of first-order diffusion is the relative
suppression of modes with larger $k$, \ie, the monotonic smoothing of
any initial density profile.

It is possible for experiments to give an upper bound for $\D$.  If one
sees no structure in the number density profile for rapidity separations
up to $\Delta\eta$ then one can conclude either that structures on such
scales were not present in the initial state, or that they were present
and were wiped out by diffusion. The time available for diffusion to act
is bounded by the freezeout epoch, $\tau_f$. Hence, one gets the limit
$\D \le \tau_f \sinh^2 \Delta \eta$.  One should be careful about this
upper limit for $\D$, if $\Delta\eta$ is very small, \ie, if one observes
fluctuations in the number density profile at all scales.  There is a
sum rule between $\D$ and the relaxation time for diffusive processes,
$\tau_R$ \cite{kelly, sgupta}: $\D=c_s^2\tau_R$.  When freezout occurs
away from a critical point, a vanishing $\D$ implies a vanishing $\tau_R$,
whereas we expect that $\tau_R$ is not zero. Hence, if we see fluctuations
of number densities at all scales $\Delta\eta$, then it is likely that
the first order theory breaks down. To find what could happen then,
we next investigate the second order theory.

At second order, Fick's law, eq.\ (\ref{fick}), can be replaced by the
form \cite{kelly,sgupta}
\beq
   0= \left(1+\tau_R \partial_t\right) n^i + {\D}\partial^i n,
\label{secondfick}\eeq
where $\tau_R$ is a relaxation time for number changing processes. This
equation is written in the rest frame of the fluid. One uses the projector
orthogonal to $u$ to write this covariantly. The covariant equation is
\beq
   0= \nu v^\mu +\tau_R \Delta^{\mu\nu} Dn_\nu
         - {\D}(d^\mu n - u^\mu Dn).
\label{secondcov}\eeq
Projecting parallel to $u$ gives 0 identically. Projecting parallel
to $v$ and using the definition $Du_\mu=Sv_\mu$ and the identity
$v_\mu Dv^\mu=0$ gives
\beq
   \nu + \tau_R D\nu = - {\D}{\widetilde D}n - S \tau_R n.
\label{secondscal}\eeq
This replaces the constraint (eq.\ \ref{scalfick}) for $\nu$ obtained
in the first order formalism.  Kelly makes the identification
${\D}/\tau_R=c_s^2$, and we retain this intuition in this work. The
quantity $\tau_R$ is currently unknown; it could be of the order of
typical QCD scales, \ie, 1 fm, or it could be significantly smaller in
AdS/QCD scenarios, for example, 0.1 fm.

In the background of boost-invariant longitudinal flows, eqs.\ (\ref{conti})
and (\ref{secondscal}) reduce to
\beq
   0 = \frac{\partial n}{\partial\tau} + \frac n\tau
         + \frac1\tau \frac{\partial\nu}{\partial\eta},\qquad{\rm and}\qquad
   0 = \tau_R \frac{\partial\nu}{\partial\tau} + \nu
         + \frac{\D}\tau \frac{\partial n}{\partial\eta}.
\label{longdiss}\eeq
This is the form that Kelly's second order diffusion equation \cite{kelly}
takes in a boost-invariant geometry.  In general one expects the
diffusion constant $\D$ and the relaxation time $\tau_R$ to depend on
the temperature, $T$. Since the time-evolution of $T$ is obtained by
solving the remaining hydrodynamic equations, one gets an explicit time
dependence of ${\D}$ and $\tau_R$, and hence one can solve the equations
by Fourier transformation. The evolution equations for the Fourier modes
can be written in the form
\beq
   \frac{\partial}{\partial\tau} \left(\matrix{n(\tau,k)\cr \nu(\tau,k)}\right)
    = -\left(\matrix{1/\tau & ik/\tau\cr
               ic_s^2k/\tau & 1/\tau_R}\right)
             \left(\matrix{n(\tau,k)\cr \nu(\tau,k)}\right).
\label{fourdiss}\eeq
Since $c_s^2$ varies within a bounded region, $c_s^2/\tau$ falls with
$\tau$, and at late times $\nu$ decays as $\exp(-\tau/\tau_R)$, and $n$
decays as $1/\tau$, thus reducing to Bjorken attenuation at late times.
The $k=0$ mode exhibits Bjorken attenuation at all times.

For more detailed analysis it is useful to
change variables to $\t=\log(\tau/\tau_R)$. The equations then become
\beq
   \frac{\partial}{\partial\t} \left(\matrix{n(\t,k)\cr \nu(\t,k)}\right)
    =  -M \left(\matrix{n(\t,k)\cr \nu(\t,k)}\right),\qquad
   M = \left(\matrix{1 & ik\cr ic_s^2k & {\rm e}^{\t}}\right).
\label{fourdissp}\eeq
Note three regions of Fourier modes---
\begin{enumerate}
\item When $k$ is sufficiently smaller than unity, the off-diagonal terms
 may be neglected to a good approximation, and the problem decouples. The
 equation for $n(\t,k)$ is then similar to the ideal case. We do not discuss
 the $k=0$ mode in the following.
\item When $k$ is sufficiently larger than $\exp(\t)$, the problem simplifies
 again. However, the hydrodynamic equations are an approximation to microscopic
 physics, and are valid for long wavelength spatial fluctuations and for
 long-time phenomena. The time scales are $\exp\t>1$, and the limit of
 $k\to\infty$ must be discarded, although the equations simplify.
\item In the remaining region the equations are fully coupled and a more
 detailed analysis is called for. This follows.
\end{enumerate}

Since $M$ depends explicitly on $\t$, a complete solution to the
differential equations cannot be obtained by just diagonalizing it. First
note that when $k\ne0$, then for either $\t\ne0$ or $c_s^2\ne1$, one
has $[M,M^\dag]\ne0$, \ie, $M$ is non-normal. As a result, $M$ is not
diagonalized by a unitary transformation, and the eigenvectors are not
orthogonal. For non-normal matrices the concept of pseudospectra can
yield powerful results \cite{pseudo}. In this paper, however,
we present a standard spectral analysis, since very detailed information
can be obtained by this route.

Since the trace and determinant of $M$ are real, the eigenvalues are
either both real or both complex,
\beq
   \lambda_\pm = \frac12(1+{\rm e}^\vartheta) \pm \frac12
         \sqrt{({\rm e}^\vartheta-1)^2 -4c_s^2k^2}.
\label{eigs}\eeq
For a given $\vartheta$, the eigenvalues are real for $|k|\le ({\rm e}^\vartheta-1) /
(2c_s)$. When they are real, then they are both positive since the first
term of the expression in eq.\ (\ref{eigs}) is greater than ${\rm e}^\vartheta/2$
and the second term is less than ${\rm e}^\vartheta/2$. When both are complex, then
the real parts are positive, since the trace is so. This implies that
$|n|^2$ and $|\nu|^2$ both must decay at long times. However, there may
be transient growth.

\begin{figure}[htb]\begin{center}
   \scalebox{0.65}{\includegraphics{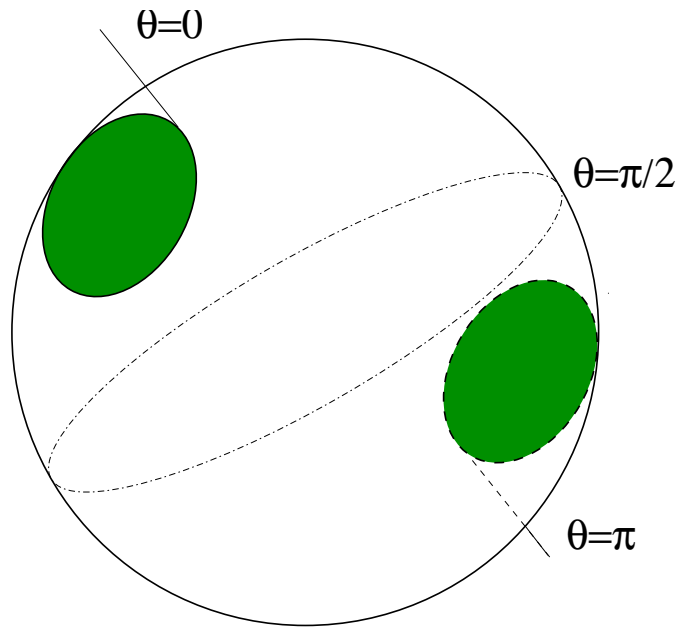}}
   \end{center}
  \caption{The parts of the sphere in which the power spectrum is
  amplified are indicated by the green patches. They are symmetrically
  placed at the poles and are bounded in the region $0\le\theta\le
  \tan^{-1}k$ and symmetrically in $\pi+\tan^{-1}(-k)\le\theta\le\pi$.}
\label{fg.transient_area}\end{figure}

One way to show that transient
growth may occur in general is to write down the evolution equation for
the power spectrum
\beq
   P(\t,k) = |n(\t,k)|^2 = {\x}^\dag A {\x} \quad{\rm where}\quad {\x} =
       \left(\matrix{n\cr\nu}\right)
     \quad{\rm and}\quad A=\left(\matrix{1 & 0\cr 0 & 0}\right).
\label{spec}\eeq
The evolution equation is
\beq
   \frac{dP(\t,k)}{d\t} = \frac{d{\x}^\dag}{d\t} A{\x}
              + {\x}^\dag A \frac{d{\x}}{d\t}
      = -{\x}^\dag{\cal M}{\x},
     \quad{\rm where}\quad {\cal M}=\left(\matrix{2 & ik\cr -ik & 0}\right),
\label{evolp}\eeq
independent of $c_s^2$ and $\t$. The quantity $-{\x}^\dag{\cal
M}{\x}/|x|^2$ is called the numerical range of the matrix $-{\cal M}$, and
is bounded by the eigenvalues of $-\cal M$, since the matrix is Hermitean.
Since the determinant is negative, one of the eigenvalues is positive and
the other negative. As a result, the numerical range has indefinite sign,
and amplification of the power spectrum is possible. The vectors $\x$ may be
parametrized as ${\x}^\dag = x(\sin\theta {\rm e}^{-i\phi}, \cos\theta)$,
where $\theta$ and $\phi$ range over the sphere. Then one has
\beq
   -{\x}^\dag{\cal M}{\x} = - 2|x|^2\sin\theta\left(\sin\theta
     + k\cos\theta\sin\phi\right).
\label{rhs}\eeq
At the equator, $\theta=\pi/2$, this quantity is negative.  There are
zeroes at the poles, $\theta=0$ and $\pi$. Other zeroes occur along the
curves $\tan\theta=-k\sin\phi$. This equation describes two closed curves
on the sphere each of which passes through one of the poles (see Figure
\ref{fg.transient_area}). In the area of the sphere inside these curves
(not containing the equator) the function is positive. Amplification
of $P(\t,k)$ can take place whenever $\x$ passes through the non-empty
region where ${\x}^\dag{\cal M}{\x}<0$.

Since initial conditions change from event to event, the probability of
transient growth is measured by the fractional area of the sphere
where transient growth may occur. The larger
the value of $k$, the closer does the zero curve come to the equator,
and hence the larger the probability of transient growth becomes. However,
large $k$ corresponds to smaller $\Delta\eta$, and hence to short range
structures in rapidity. Since such structures are efficiently erased in
Fick diffusion, their observation in a significant fraction
of events would serve as a signal of Kelly diffusion.

\begin{figure}[htb]\begin{center}
   \scalebox{0.55}{\includegraphics{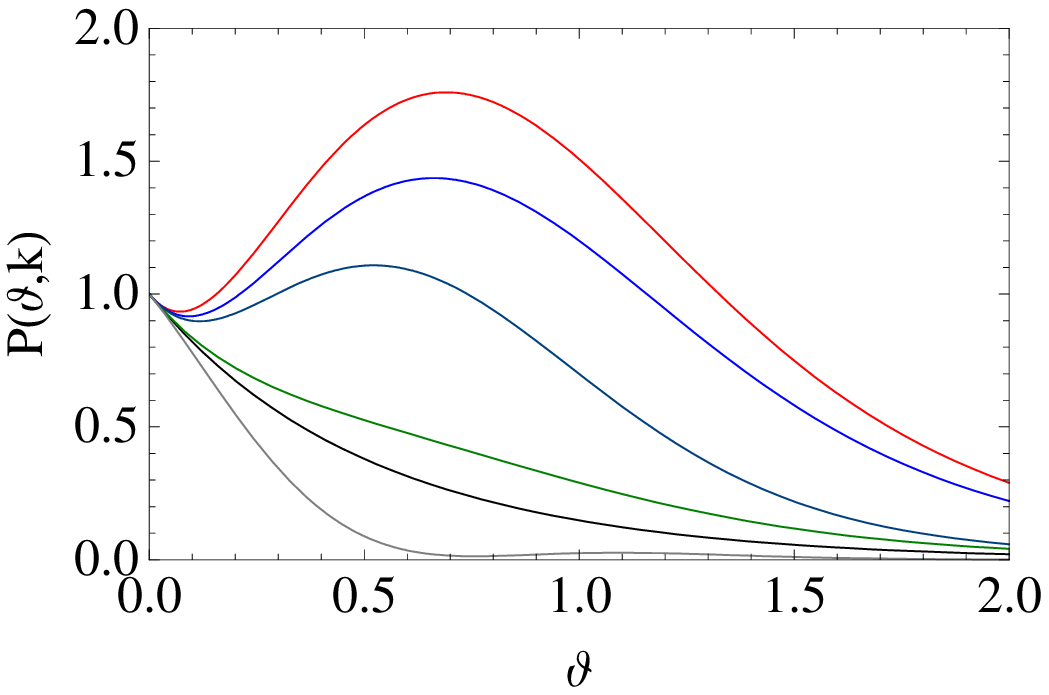}}\hfil
   \scalebox{0.55}{\includegraphics{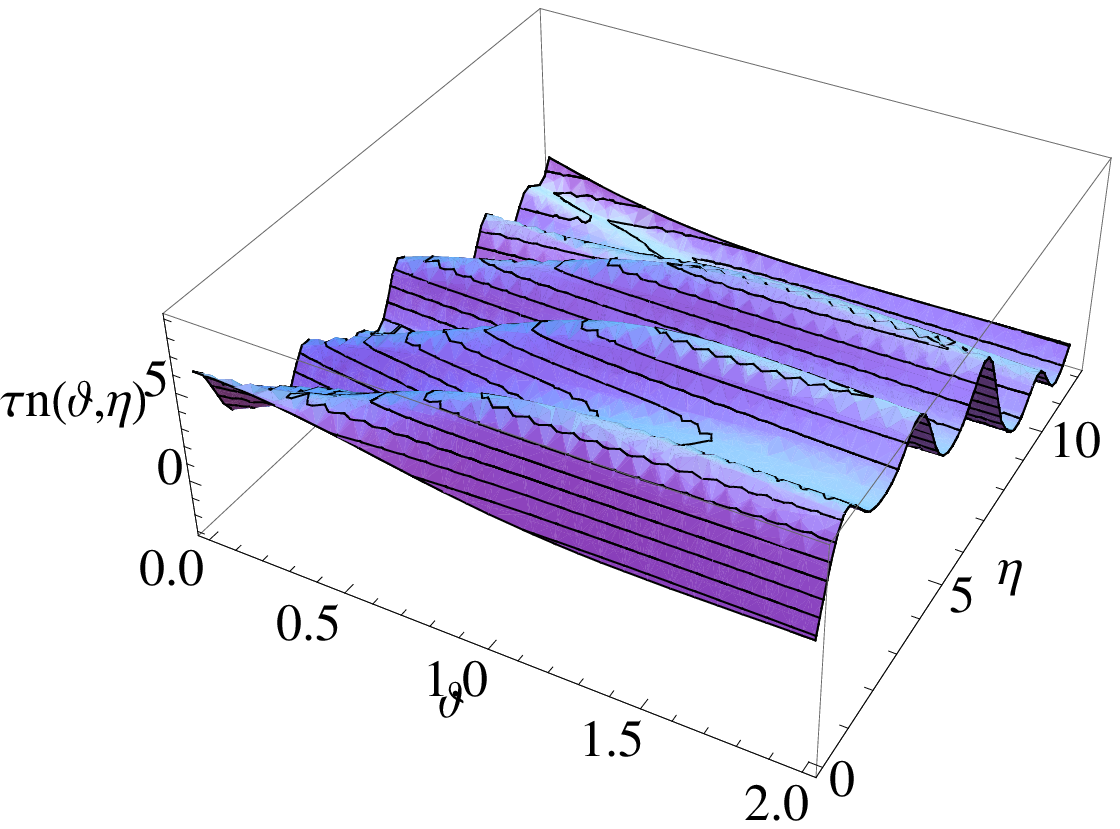}}
   \end{center}
  \caption{The evolution of the power spectrum, $P(\t,k)$, and the number
  profile, $\tau n(\t,\eta)$, for initial conditions containing only a
  discrete set of non-vanishing modes $k=1/8$ (black), $1/4$ (red),
  $1/2$ (green), 1 (blue), 2 (dark blue) and 4 (gray), which are taken
  with a flat power spectrum at $t=0$. We have chosen $c_s^2=1/3$ for
  this example. The initial conditions on the corresponding $\nu(0,k)$
  are assigned at random. The evolution of the power spectrum is shown
  on the left and of the profile on the right. In contrast to the
  intuition from the Fick theory, the profile may steepen at intermediate
  times.}
\label{fg.kellyresult}\end{figure}

\begin{figure}[hbt]\begin{center}
   \scalebox{0.55}{\includegraphics{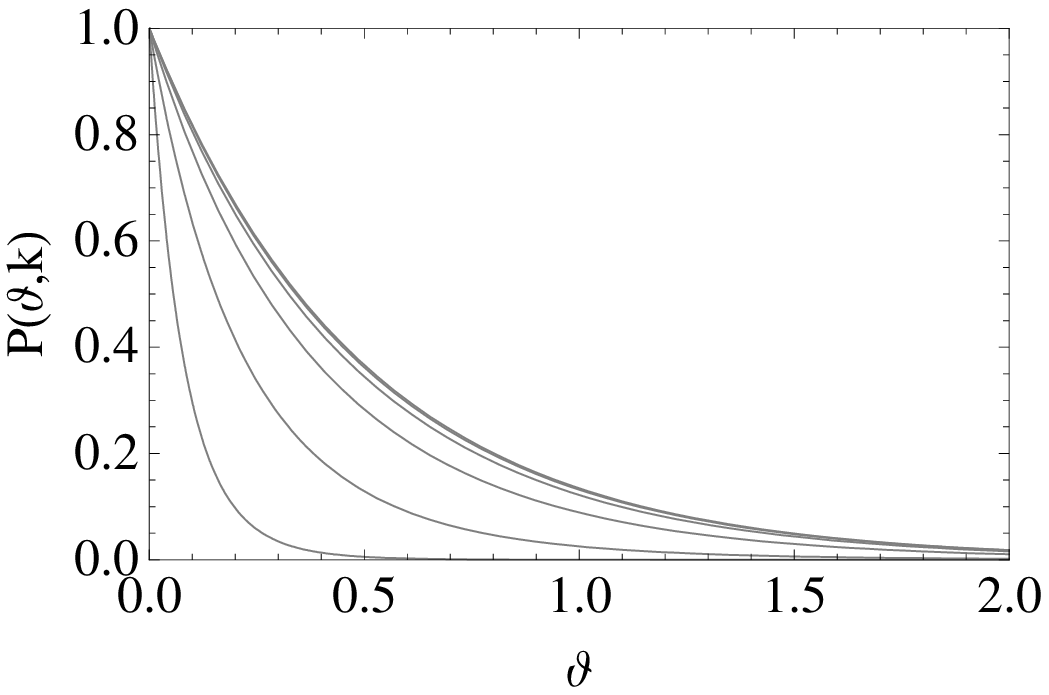}}\hfil
   \scalebox{0.55}{\includegraphics{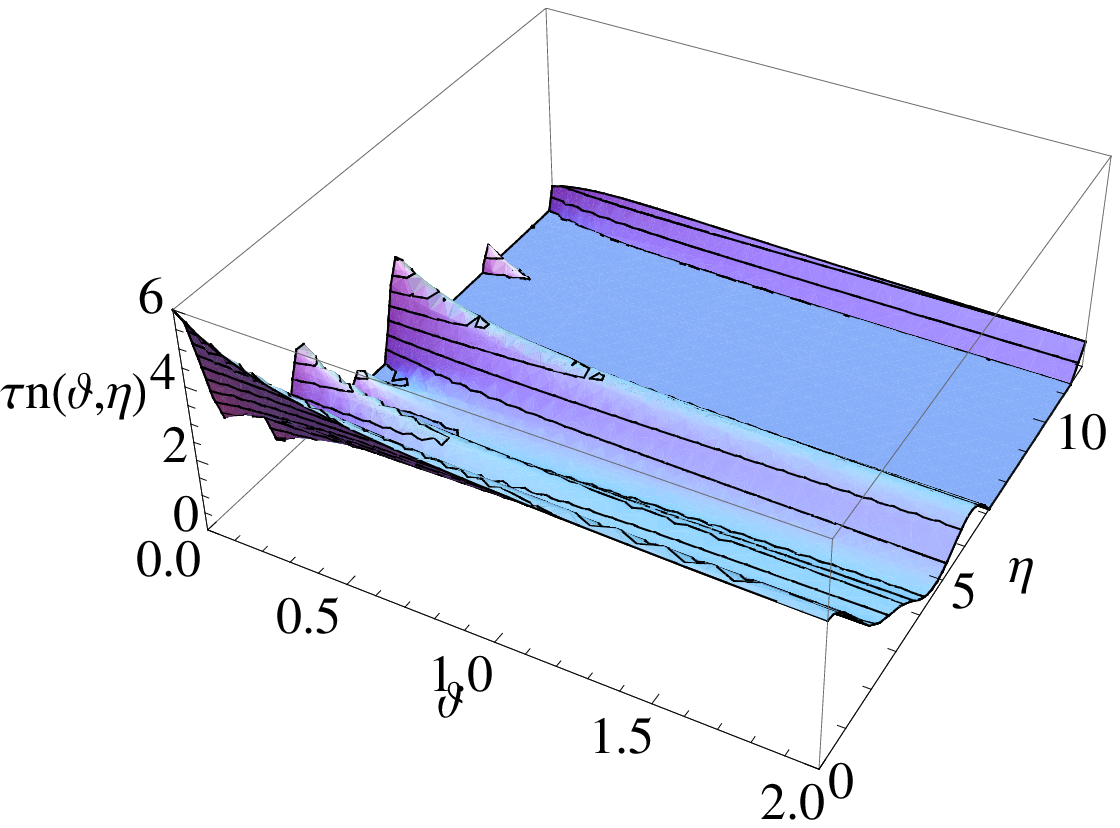}}
   \end{center}
  \caption{Same as Figure \ref{fg.kellyresult}, but using the first
   order diffusion equation. The evolution of the power spectrum is
   governed by eq.\ (\ref{solvdiff}).}
\label{fg.fickresult}\end{figure}

In Figure \ref{fg.kellyresult} we show an example of the evolution of
number density profiles using eq.\ (\ref{fourdissp}),
starting from random initial conditions. In terms of Fourier modes it
is clear that transient growth may occur for any $k$. The typical time
scale of this growth is $\t\approx1$, \ie, $\tau\approx e\tau_R$. However,
even at $\t\approx2$, \ie, $\tau\approx e^2\tau_R$ the evolution does
not begin to resemble Bjorken attenuation. If $\tau_R$ is in the range
of 1/2 to 2 fm, then the evolution of number densities may be 
dominated by transients.  Such long time scales are good news, because
they allow one to extract information on the early stages of the fireball.

Figure \ref{fg.fickresult} shows the evolution in the Fick theory
of $P(\t,k)$ and $\tau n(\t,k)$ starting from the same initial conditions as above.
The difference between first order (Fick) and second order (Kelly)
theories of diffusion seems to be fairly fundamental. The main intuitive
understanding of diffusion from the Fick theory is the following: if
initial conditions set up sharp density gradients, then diffusion always
smooths these out on all scales, monotonically, \ie, as time evolves the
gradients get smoother and smoother. The Kelly theory can violate this
intuition. Depending on initial conditions, gradients may be transiently
amplified, at different scales at different times. The asymptotic final
state is the same in both theories. However, since freezeout may not
occur at asymptotically late times, the number density profiles at
freezeout could be different. Some feel of the differences between the
theories can be obtained by comparing Figures \ref{fg.kellyresult} and
\ref{fg.fickresult}. Note, in particular, that the solution of Kelly's
equation contains structure in the number density profile at scales of
$\Delta\eta\simeq1$ even at times $\tau\simeq7.5\tau_R$.  In contrast,
for the solution of Fick's equation, the scales of $\Delta\eta$ on
which structures are seen are much larger.

\begin{figure}[hbt]\begin{center}
   \scalebox{0.75}{\includegraphics{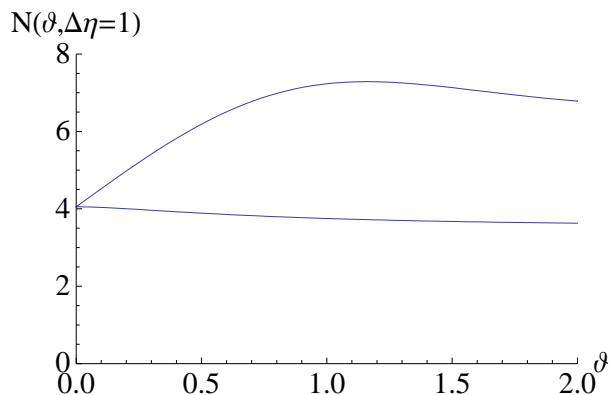}}
   \end{center}
  \caption{The total conserved charge, $N$, within a window $\Delta\eta=1$
   (between $\eta=0$ and $\eta=1$) for the initial conditions used in Figures
   \ref{fg.kellyresult} and \ref{fg.fickresult}. The upper line is for the
    Kelly theory and the lower one for the Fick theory.}
\label{fg.windowedcharge}\end{figure}

In Figure \ref{fg.windowedcharge} we show the conserved charge,
$N(\t,\Delta\eta=1)$, in a rapidity window $\Delta\eta=1$ (between
$\eta=0$ and $\eta=1$) as a function of $\t$. In both theories there
is a tendency for the charge to become independent of $\t$ at late
times. This happens because diffusion must eventually wipe out all
spatial structure in $n(\t,\eta)$, and once the profile has flattened,
the number density is locally conserved.  The differences between the two
theories are due precisely to the transients that we have studied. In the
Fick theory transient lifetimes are extremely small even for $k<1$,
hence the asymptotic behaviour sets in very early. This can also be seen
from the profiles in Figure \ref{fg.fickresult}. In the Kelly theory, on
the other hand, transients could be long lasting (depending on initial
conditions). This is seen in the evolution of the density profiles
in Figure \ref{fg.kellyresult}, as well as of $N(\t,\Delta\eta=1)$
(Figure \ref{fg.windowedcharge}).

\begin{figure}[bth]\begin{center}
   \scalebox{0.5}{\includegraphics{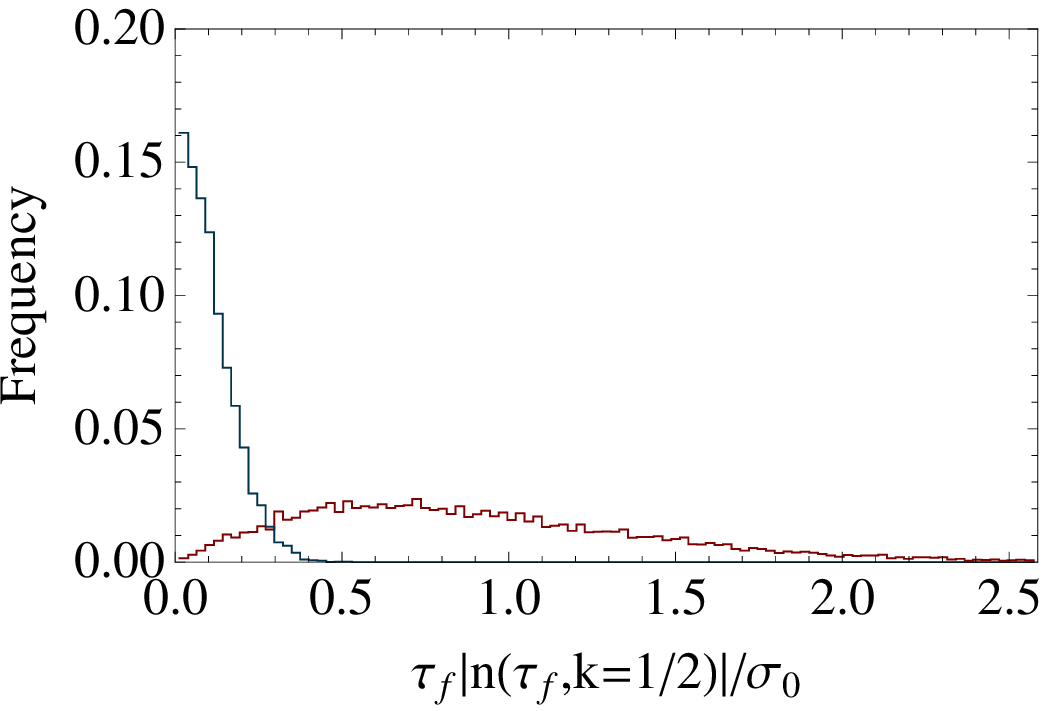}}
   \scalebox{0.5}{\includegraphics{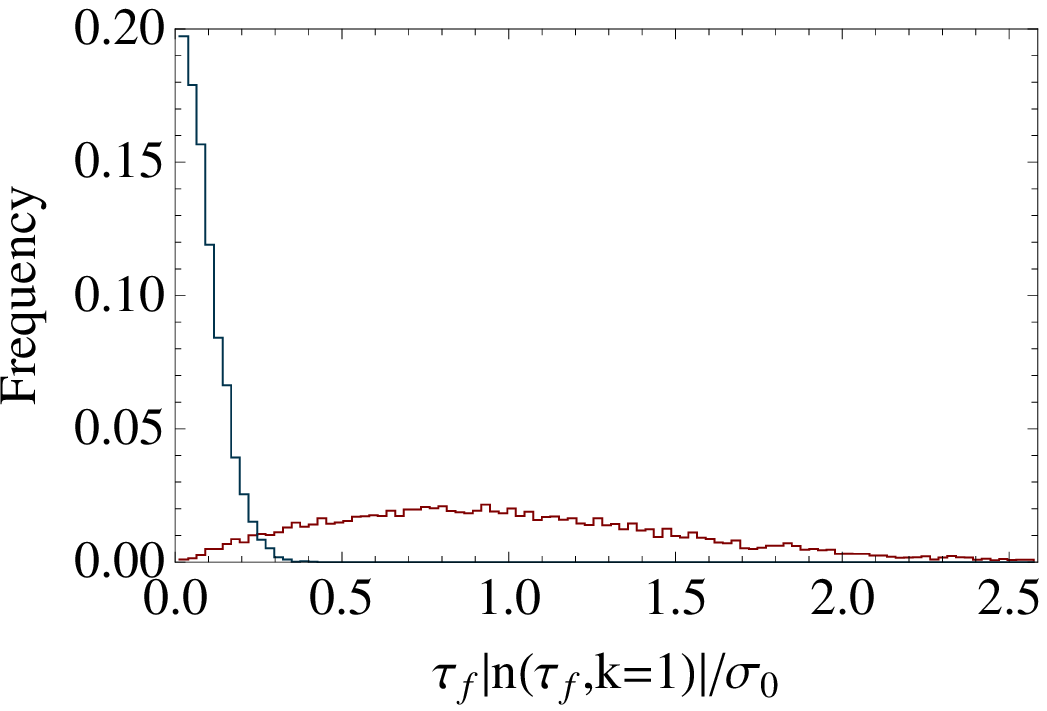}}
   \scalebox{0.5}{\includegraphics{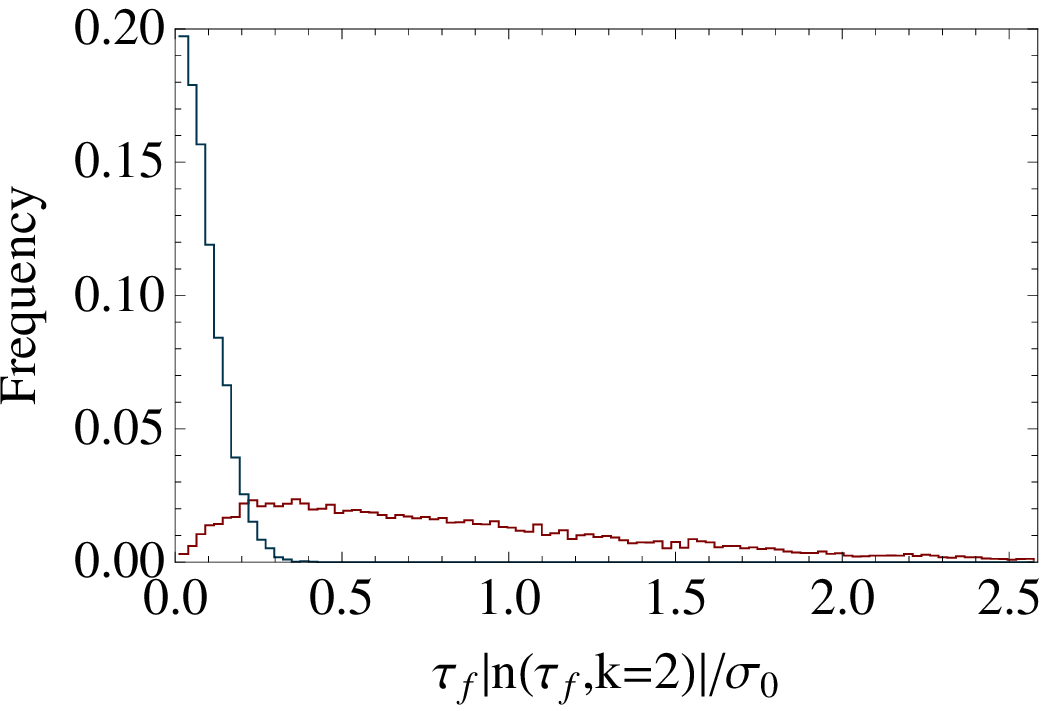}}
   \end{center}
  \caption{The distributions of the square root of the power spectrum of
    eq.\ (\ref{powch}) in the Fick (blue) and the Kelly (red) theories
    of diffusion for various $k$ at $\tau_f$ corresponding to $\t=2$.
    The results are from Monte Carlo simulations with 10000 points for each
    simulation. Initial conditions were chosen from a Gaussian distribution
    with zero mean and unit variance. All other parameters correspond to
    the values used to obtain the results in Figures \ref{fg.kellyresult}
    and \ref{fg.fickresult}. The frequency histograms are normalized to
    have unit area.}
\label{fg.distrib}\end{figure}

Our discussion till now has been geared towards identifying the
fundamental differences between the Fick and Kelly theories of
diffusion. However, the observables that we have discussed may not be
the best suited for heavy-ion physics. This is because we have looked at
the consequences of starting from a given initial condition. The initial
conditions for the fireball change from event to event, and averages
over events are unlikely to show overdense or underdense regions.
At the LHC particle multiplicities are expected to be high enough that
one could build up the profile function $n(\tau_f,\eta)$ at the freezeout
epoch, $\tau_f$,  in a single event. Then by observation of a relatively
small number of events, one could distinguish between Kelly and Fick
diffusion. However, other techniques are required when the particle
multiplicities are smaller.

It would be simplest to construct observables that depend on the power
spectrum at freezeout. In an event with $N_t$ tracks, one has the charge
$q_j$ of the $j$-th track and its rapidity $\eta_j$. Given these, one
may construct the power spectrum of the charge
\beq
   \P(\tau_f,k) = \left|\sum_{j=1}^{N_t} q_j \exp(-ik\eta_j)\right|^2.
\label{powch}\eeq
Since the $q_j$ are conserved numbers and not their densities, one has
$\P(\tau_f,k)=\tau_f^2 P(\tau_f,k)$.  One may similarly construct the
power spectrum of any conserved quantity: $S$, $B$ or even the proton
number, by replacing $q_j$ in the above formula by the quantum number
under discussion. Again, a plot of $\P(\tau_f,k)$, averaged over events,
as a function of $k$ will smear out many of the effects that we have
discussed. Some information can be gained by comparing the results with
those obtained from `mixed events', \ie, when tracks from different
events are randomly thrown together into an artificial event with the
right multiplicity.

We find that the distinction between Kelly and Fick theories can
be observed most directly in the event-to-event distribution of
$\tau_f|n(\tau_f,k)| \propto \sqrt{\P(\tau_f,k)}$. The proportionality
constant requires knowledge of the initial volume. However, both the
first and second order diffusion equations being linear, the overall
normalization of the $n$ and $\nu$ are immaterial. We take advantage
of this to work with the dimensionless quantities $\sqrt{\P(\tau_f,k)}
= \tau n(\tau,k)/\sigma_0$ and $\tau \nu(\tau,k)/\sigma_0$, where the
dimensional quantity $\sigma_0$ need not be specified except when making
an actual connection to models of the initial state. Even so, a prediction
of the distribution of $\sqrt{\P(\tau_f,k)}$ is not possible since
the initial distribution is not known. However, starting from the same
initial distribution the two theories give rise to completely different
distributions at freezeout, and the differences can be investigated.

An example is shown in Figure \ref{fg.distrib}. Here we started with
initial conditions drawn from a Gaussian distribution with zero mean
and unit variance. Then from eq.\ (\ref{solvdiff}) it is
clear that for Fick diffusion, the distribution of $\sqrt{\P(\tau_f,k)}$
at freezeout is also a Gaussian, with standard deviation $\exp[-\D
k^2 (1/\tau_0 - 1/\tau_f )]$.  Note the strong dependence of the
variance on $k$, which is also apparent in Figure \ref{fg.distrib}.
The distribution of this quantity in the Kelly theory cannot be derived
so easily. We determined it through a Monte Carlo simulation, deriving
the distribution at freezeout time by evolving many different samples
of initial conditions. The resulting distribution is clearly different.
It does not peak at zero, the position of the peak shifts with $k$, and
it has a very long tail. These features are intimately connected with the
transient amplification phenomenon which has been discussed above. Beyond
the range shown in the figures, as $k$ decreases, the frequency of small
$\sqrt{\P}$ increase, the tail shrinks, and the distinction between Fick
and Kelly theories is lost in the limit of $k\to0$. Most importantly,
the fact that in all theories the long-time behaviour is dominated by
Bjorken attenuation means that the distribution of $\P(\tau_f,k)$ (or its
square root) is independent of $\tau_f$. Event-to-event distributions of
this kind therefore seem to be the most promising observable distinction
between first and second order diffusion.

One of the most widely studied signals of a critical end-point of QCD
is the distribution of event-to-event fluctuations of conserved
quantities contained within a rapidity acceptance window. The present work,
with its removable simplifications, shows that these measures could be
influenced, possibly strongly, by the nature of the transport process---
whether first or second order. Therefore our understanding of the nature
of diffusion needs to be improved before event-to-event fluctuations
can be interpreted. We have shown that one way to do this is to study
the event-to-event distribution of the power spectrum of the conserved
charge (see eq.\ \ref{powch}), at $\tau_f$ for different $k$ (see
Figure \ref{fg.distrib}). There are qualitative differences between the
results for the two theories of diffusion.

In summary, here are our main conclusions---
\begin{enumerate}
\item The Fourier coefficients, $n(\tau,k)$, and the power spectrum,
 $P(\tau,k)$, of the number density profile (see eq.\ \ref{fourier}),
 obtained by Fourier transforming in the rapidity (which is equal to
 $\eta$ for boost invariant flow) are of interest in the study of
 diffusion and hydrodynamics. One can construct number density profiles
 for any quantum number which is conserved in strong interactions. The
 obvious ones are $B$ (or $N_p$), $Q$, and $S$. A power spectrum can
 also be constructed from experimental data (see eq.\ \ref{powch}), and 
 easily compared to theory.
\item The net conserved number is obviously independent of the kind of
 hydrodynamics. As a result, the event-to-event distribution of this quantity
 tells us directly about initial conditions. The net charge within a given
 rapidity window, however, may evolve with the hydrodynamics (see Figure
 \ref{fg.windowedcharge}). Since freezeout may occur at different times
 for different centrality, this may show up as a dependence of the net
 observed charge on centrality.
\item The Fourier modes give interesting information on the nature
 of the transport theory. In Kelly (second order) diffusion, the power in
 short range modes, \ie, modes with large $k$, could be large. This is
 impossible in Fick (first order) hydrodynamics. The simplest of observables,
 \ie, a plot of $\P(\tau_f,k)=\tau_f^2|n(\tau_f,k)|^2$ as a function of $k$,
 for an event, can potentially distinguish between these two theories. One
 thing to note is that this must be done on an event-to-event basis. Averaging
 over events before constructing the power spectrum could wash out the signal.
\item Event-to-event distributions of the observed power spectrum (see
 eq.\ \ref{powch}), or its square root, for one or more values of $k$ can
 potentially distinguish between Fick and Kelly theories of diffusion (see
 Figure \ref{fg.distrib} and the corresponding discussion in the text).
\item If the Kelly theory is ruled out by such observables, then one can
 construct an upper bound on the transport coefficient,
 \beq
    \D\le {\tau_f}{\sinh^2\Delta\eta},
 \label{estimate}\eeq
 using the rapidity interval $\Delta\eta$ over which all structure has been
 washed out and an independent estimate of $\tau_f$.
\end{enumerate}

This work was motivated by discussions at the INT program ``The QCD
Critical Point''. SG would like to thank the participants, especially
V.\ Koch, G.\ Roland, M.\ Stephanov and N.\ Xu, for discussions, and
gratefully acknowledge the hospitality of the Institute of Nuclear Theory
at the University of Washington during this period.

\end{document}